\documentstyle[12pt]{article}

0
   
\begin{document}

%define commands
\newcommand{\MqV}{$\cal M$$_{q}$($\cal V$)}
\def\II{\relax{\rm 1\kern-.35em1}}
\def\IP{\relax{\rm I\kern-.18em P}}
\renewcommand{\theequation}{\thesection.\arabic{equation}}
\csname @addtoreset\endcsname{equation}{section}

\hspace{12cm} hep-th/9701150

\vspace{10 mm}

\begin{center}

{}~\vfill

{\large \bf M and F-Theory Instantons, $N\!=\!1$ Supersymmetry and}  

\end{center}
  
\vspace{1 mm}

\begin{center}

{\large \bf Fractional Topological Charge}
  
\end{center}

\vspace{20 mm}

\begin{center}

{\bf C\'{e}sar G\'{o}mez and Rafael Hern\'{a}ndez} 
  
\vspace{10 mm}

{\em Instituto de Matem\'{a}ticas y F\'{\i}sica Fundamental,
CSIC, \protect \\ Serrano 123, 28006 Madrid, Spain}  
  
\vspace{40 mm}   
  
\end{center}      

\begin{abstract}
We analize instanton generated superpotentials for three dimensional 
$N\!=\!2$ supersymmetric gauge theories obtained by compactifying on 
$S^1$ $N\!=\!1$ four dimensional theories. For $SU(2)$ with $N_f=1$, we
find 
that the vacua in the decompactification limit is given by the singular 
points of the Coulomb branch of the $N\!=\!2$ four dimensional theory (we also 
consider the massive case). The 
decompactification limit of the superpotential for pure gauge theories 
without chiral matter is interpreted in terms of `t Hooft's fractional 
instanton amplitudes.
\end{abstract}

\pagebreak

%%%%%%%%%%%%%%%%%%%%%%%%%%%%%%%%%%%%%%%%%%%%%%%%%%%%%%%%%
%%%%%%%%%%%%%%%%%%%%%%%%%%%%%%%%%%%%%%%%%%%%%%%%%%%%%%%%%

\section{Introduction.}

Some of the deepest dynamical problems in gauge theories, such as the 
confinement problem, appear as tractable issues once we pass from four to 
three dimensions \cite{P}. Recently, the study of $N\!=\!2$ supersymmetry in 
three dimensions has began to shed some light on the more difficult 
dynamics of four dimensional $N\!=\!1$ theories \cite{SW3d}. Moreover, the 
interplay between $N\!=\!2$ in three dimensions and $N\!=\!4$ in four 
dimensions is the one existing between M and F-theory compactifications on 
Calabi-Yau fourfolds.
  
Using elliptically fibered Calabi-Yau fourfolds, the Coulomb branch of an 
ADE $N\!=\!2$ gauge theory in three dimensions can be defined by means 
of the resolution of the corresponding ADE singularity \cite{KV}. Instantons 
defined in terms of aritmetic genus one divisors \cite{Wsp} of the resolved 
fourfold provide an $R$-dependent superpotential \cite{KV}, with 
$R$ determined by the class of the elliptic fiber \cite{Wsp}. The 
$R \rightarrow 0$ limit reproduces the known results in $N\!=\!2$ three 
dimensional theories \cite{AHW}, while the $R \rightarrow \infty$ limit 
defines a superpotential in four dimensions, with $N\!=\!1$ 
supersymmetry, compatible with the $\hbox{tr} (-1)^F$ computation \cite{Wind}, 
and the cluster derivation of gaugino condensates \cite{SVZ,A}. The main 
interest of this four dimensional limit is that it can not be trivially described 
directly in four dimensions or, equivalently, in the F-theory compactification 
on the fourfold, where the absence of Coulomb branch forbids the resolution 
of singularities. Thus, the $R \rightarrow \infty$ limit provides some 
information on the strong infrared dynamics taking place in the confinement 
regime of the uncompactified theory.
  
In this letter we study the case of $SU(2)$ with $1$ flavor. For 
$1$ flavor, we find an $R$-dependent superpotential on the Coulomb branch 
of $N\!=\!2$ in three dimensions, which in the $R \rightarrow \infty$ limit 
provides a set of minima in agreement with the result arising from soft 
breaking ($N\!=\!2$ to $N\!=\!1$) the exact solution \cite{SW2} for the 
Coulomb branch of four dimensional $N\!=\!2$ with $N_f=1$. 
The other issue we consider is the direct 
interpretation of the $R \rightarrow \infty$ limit of the pure $N\!=\!1$ 
four dimensional superpotential in terms of `t Hooft \cite{tHcmp} fractional 
instanton amplitudes \cite{CG}. In fact, for $SU(N_C)$ gauge theories, the topology 
of the residual gauge transformations compatible with the set of twisted boundary 
conditions representing the non-vanishing ${\bf Z}_{N_C}$-magnetic flux 
\cite{tHtw}, boundary conditions compatible with the confinement phase, 
reproduce in the four dimensions context the Dynkin structure \cite{KV} of 
the singularity resolution used in the definition of the Coulomb branch 
of three dimensional $N\!=\!2$.

%%%%%%%%%%%%%%%%%%%%%%%%%%%%%%%%%%%%%%%%%%%%%%%%%%%%%%%%%%%%%%%%%%%%%
%%%%%%%%%%%%%%%%%%%%%%%%%%%%%%%%%%%%%%%%%%%%%%%%%%%%%%%%%%%%%%%%%%%%%

\section{M-theory Instantons.}

We will consider M-theory compactifications on a Calabi-Yau fourfold $X$,  
which lead to three dimensional $N\!=\!2$ supersymmetry. If $X$ admits an 
elliptic fibration,
\begin{equation}
E \longrightarrow X \stackrel{\Pi}{\longrightarrow} B,
\end{equation}
we can define F-theory compactifications on $X$ which lead to four dimensional 
$N\!=\!1$ supersymmetry. If we assume that on a codimension one locus $C 
\subset B$ the elliptic fiber is degenerate, of ADE type in Kodaira's 
classification, then we obtain an $N\!=\!1$ ADE-gauge theory in four dimensions. 
This $N\!=\!1$ four dimensional theory results from compactifying on $C$ the 
$7$-brane worldvolume. The bare coupling constant is then given by \cite{KV}
\begin{equation}
\frac {1}{g_4^2} = V_C.
\end{equation}
Through further compactification on $S^1$, we recover the $N\!=\!2$ three 
dimensional theory defined by the M-theory compactification on $X$.
  
Denoting by $\epsilon$ the class of the elliptic fiber, $E$, we can relate, 
by a chain of dualities, M-theory compactified on $X$ with type II$_B$ 
compactifications on $B \times S^1$, where the radius of $S^1$ scales 
like $\frac {1}{\epsilon}$. The decompactification limit $\epsilon 
\rightarrow 0$ ($R \rightarrow \infty$) corresponds to the F-theory 
compactification, while the $\epsilon \rightarrow \infty$ 
($R \rightarrow 0$) limit does correspond to M-theory compactification.
  
As it is well known \cite{SW3d}, $N\!=\!2$ supersymmetric three dimensional 
pure gauge theories posses a classical Coulomb branch of dimension 
equal to the rank $r$ of the gauge group. This Coulomb branch is parametrized
, if we define the theory by compactifying the $N\!=\!1$ theory in four dimensions, in
terms of the Wilson line in the internal direction. At the classical level we have a set of
$r+1$ singular points corresponding to Wilson lines in the center of the group. 
If fundamental fermions
are absent, then these point represent classical restoration of non abelian 
symmetry (see some comments on this issue in the last section).
 In three dimensions, instantons 
are characterized by their monopole magnetic charge. They generate non 
perturbatively a superpotential \cite {AHW} of the type
\begin{equation}
W = \sum _{i=1}^{r} \hbox{exp} (- \Phi_i),
\label{eq:3}
\end{equation}
with $\Phi_i$ the $r$ complex scalars used to parametrize the Coulomb 
branch. The existence of this superpotential is mainly due to the fact that instantons 
in three dimensions have only two fermionic zero modes, as superconformal 
invariance is absent.
  
The M-theory origin of the superpotential (\ref{eq:3}) has recently 
been presented in references \cite{Wsp,KV}. In fact, instantons in 
M-theory are defined \cite{Wsp} wrapping the euclidean $5$-brane on a $6$-cycle $D$, 
contained in $X$, satisfying 
\begin{equation}
\chi({\cal O_D})=1-h^{3,0}-h^{2,0}+h^{1,0}=1.
\label{eq:4}
\end{equation}
Condition (\ref{eq:4}) is equivalent to the existence, in uncompactified 
spacetime, of two fermionic zero modes, which implies the generation of a superpotential. 
  
When $X$ admits an elliptic fibration, vertical instantons, defined by 
a divisor $D$ satisfying (\ref{eq:4}), and such that $\Pi(D)$ is 
codimension one in $B$, survive in the uncompactified F-theory limit defined 
as $\epsilon \rightarrow 0$ \cite{Wsp}. In fact, through a chain of dualities the 
$5$-brane instanton can be interpreted in the type II$_B$ language as a 
$3$-brane wrapping the four cycle $\Pi(D)$.
  
Let us now consider a divisor $D$ with $\Pi(D)=C \subset B$, with $C$ the 
codimension one locus in $B$ where the elliptic fiber develops an ADE 
singularity. We will also assume that $h^{1,0}(C)=h^{2,0}(C)=0$, which 
from the point of view of F-theory implies the absence of adjoint matter 
in the four dimensional theory. Using Hirzebuch-Riemann-Roch theorem, we 
get for this divisor
\begin{equation}
\chi({\cal O}_D) = C_2(G),
\label{eq:5}
\end{equation}
with $C_2(G)$ the dual Coxeter number. Clearly, this divisor $D$ does not 
contribute to the superpotential. In fact, to go to the Coulomb branch 
in three dimensions is equivalent to performing the resolution of the 
singular elliptic fiber at $C$ \cite{KV}. By this procedure, we get a set of $r+1$ 
irreducible components $E_i$ satisfying $E_i \cdot E_i = -2$, with 
intersection matrix defined by the corresponding affine Dynkin diagram. 
Each of these components can be used in order to define a divisor $D_i$, 
obtained by fibering $E_i$ on $C$, with 
\begin{equation}
\chi({\cal O}_{D_i}) = 1. 
\label{eq:6}
\end{equation}
Moreover, as pointed out in reference \cite{KV}, these divisors are constrained 
by the standard relations between the roots of a Lie algebra. 
Denoting $\alpha_i$ the 
roots of the Lie algebra, we have, for a group of rank $r$,
\begin{equation}
\sum_{i=1}^{r} a_i \alpha_{i} = \tilde{\alpha},
\end{equation}
with $\tilde{\alpha}$ the biggest root, which defines the extra point in the 
affine Dynkin diagram. Moreover (see table),
\begin{equation}
\sum _{i=1}^r a_i = C_2(G)-1.
\end{equation}

\begin{center}

\begin{tabular}{|c|c|c|}     \hline\hline
        $ {\bf A_{n-1}}$   & $\tilde{\alpha} = \sum_{i=1}^{n-1} \alpha_i$                                                                              &  $C_2(G)=n$      \\ \hline
        $ {\bf D_{n}}$     & $\tilde{\alpha} = \alpha_1 + 2 \alpha_2 + \cdots + 2 \alpha_{n-2} + \alpha_{n-1} + \alpha_n$                              &  $C_2(G)=2n-2$   \\ \hline
        $ {\bf E_{6}}$     & $\tilde{\alpha} = \alpha_1 + 2 \alpha_2 + 2 \alpha_3 + 3 \alpha_4 + 2 \alpha_5 + \alpha_6$                                &  $C_2(G)=12$     \\ \hline
        $ {\bf E_{7}}$     & $\tilde{\alpha} = 2 \alpha_1 + 2 \alpha_2 + 3 \alpha_3 + 4 \alpha_4 + 3 \alpha_5 + 2 \alpha_6 + \alpha_7$                 &  $C_2(G)=18$     \\ \hline
        $ {\bf E_{8}}$     & $\tilde{\alpha} = 2 \alpha_1 + 3 \alpha_2 + 4 \alpha_3 + 6 \alpha_4 + 5 \alpha_5 + 4 \alpha_6 + 3 \alpha_7 + 2 \alpha_8$  &  $C_2(G)=30$     \\ \hline\hline
\end{tabular}

\end{center}

A set of $r$ irreducible 
components $E_i$ can be related to the roots $\alpha_{i}$, while the extra 
one, that we will denote $E_0$, can be associated to $- \tilde{\alpha}$ \footnote{Notice 
that the extra component does not contribute to the Picard group of the 
fourfold.}, which defines the extra point of the affine Dynkin 
diagram. Thus, we get the relation
\begin{equation}
\sum_{i=1}^r a_i E_i + E_0 = E, 
\label{eq:9}
\end{equation}
with $E \cdot E =0$, the class of the elliptic fiber. The contributions of 
these divisors to the superpotential are given by 
\begin{equation}
W= \sum_{i=0}^{r} \hbox{exp} [-V(C)\cdot V(E_i)].
\label{eq:10}
\end{equation}
  
Using (\ref{eq:9}), and for $V(E) = \epsilon ( \sim \frac {1}{R})$, the 
superpotential (\ref{eq:10}) becomes \cite{KV}
\begin{equation}
W= \sum_{i=0}^{r} \hbox{exp} [-V(C)\cdot V(E_i)] + \gamma 
\hbox{exp} [ \sum_{i=1}^{r} a_i \: \: V(C) \cdot V(E_i) ],
\label{eq:11}
\end{equation}
with 
\begin{equation}
\gamma = exp \left( - \frac {1}{g_3^2 \cdot R} \right).
\label{eq:12}
\end{equation}

The main interest of (\ref{eq:11}) is that it provides us with an $R$-dependent 
superpotential, with well behaved limits both in the $N\!=\!2$ three dimensional 
case and the $N\!=\!1$ case in four dimensions. In fact, the $R \rightarrow 0$ 
limit leads to the $N\!=\!2$ superpotential (\ref{eq:3}) generated by 
instantons, and in the $R \rightarrow \infty$ limit it gives a set of 
$C_2(G)$ minima, in perfect agreement with the Witten index computation \cite{Wind}.

%%%%%%%%%%%%%%%%%%%%%%%%%%%%%%%%%%%%%%%%%%%%%%%%%%%%%%%%%%%%%%%%%%%%%
%%%%%%%%%%%%%%%%%%%%%%%%%%%%%%%%%%%%%%%%%%%%%%%%%%%%%%%%%%%%%%%%%%%%%

The superpotential (\ref{eq:11}) generalizes the one obtained in reference 
\cite{SW3d}, for the particular case of $SU(2)$, to arbitrary gauge 
groups. The derivation in \cite{SW3d} starts with the solution to the 
$N\!=\!4$ theory in three dimensions, which is given by the 
Atiyah-Hitchin \cite{AH} manifold 
\begin{equation}
y^2 = x^2 v +\gamma x, 
\end{equation}
where $v=x-u$, and $\gamma$ is the dynamically generated scale, 
in adequate units. This theory corresponds to the $R \rightarrow 0$ limit of 
the $N\!=\!2$ supersymmetric pure gauge theory on ${\bf R}^3 \times S^1$. If  
we now softly break $N\!=\!2$ to $N\!=\!1$ by the addition of a superpotential 
$\epsilon u$, the $R$ dependent superpotential for the $N\!=\!1$ theory 
becomes
\begin{equation}
W = \gamma^2 \lambda ( \tilde{y}^2 - \tilde{x}^2 v - \tilde{x}) 
+ \epsilon (\gamma \tilde{x} - v),
\end{equation}
with $\gamma \tilde{x} = x$, $\gamma \tilde{y} = y$. Now, an effective 
superpotential for $\tilde{x}$ can be written by solving 
$\frac {\partial W}{\partial \lambda} = \frac {\partial W}{\partial \tilde{y}} =
\frac {\partial W}{\partial v} = 0$,
\begin{equation}
W = \epsilon \left( \gamma \tilde{x} + \frac {1}{\tilde{x}} \right),
\end{equation}
which coincides with (\ref{eq:11}) when the identification
\cite{SW3d} 
$\frac {\epsilon}{\tilde{x}} = exp(-V(C) \cdot V(E_1))$ is made\footnote{If we take the $\epsilon \rightarrow 
\infty $ limit, we can define a blow up through the double limit $\lim_{\epsilon \rightarrow 
\infty, \tilde{x} \rightarrow \infty} 
\frac {\epsilon}{\tilde{x}} = exp(-V(C) \cdot V(E_1))$. By
interpretting the blow up parameters $V(E_1)$ in terms of $x$, we
observe that their definition implies $\tilde{x} \rightarrow
\infty$, and $\gamma \rightarrow 0$.}

%%%%%%%%%%%%%%%%%%%%%%%%%%%%%%%%%%%%%%%%%%%%%%%%%%%%%%%%%%%%%%%%%%%%%
%%%%%%%%%%%%%%%%%%%%%%%%%%%%%%%%%%%%%%%%%%%%%%%%%%%%%%%%%%%%%%%%%%%%%

\section{The QCD case.}

Let us now add flavors transforming in the fundamental representation. 
For simplicity, we will consider the case of $SU(2)$. The
$R$-dependent superpotential can be derived using the
Atiyah-Hitchin manifold for massless $N_f=1$ $N\!=\!4$
supersymmetric $SU(2)$ theory,
\begin{equation}
y^2 = x^2 v +\gamma_1.
\label{eq:31}
\end{equation}

In this case, the superpotential we obtain, following the same 
steps as in \cite{SW3d}, is
\begin{equation}
W = \epsilon \left( \gamma_1 \tilde{x} + \frac {1}{\tilde{x}^2} \right),
\label{eq:32}
\end{equation}
which for $R \rightarrow 0$, gives us the superpotential
\begin{equation}
W=\frac {\epsilon}{\tilde{x}^2}
\label{eq:33}
\end{equation}
  
The minima, in the $R \rightarrow \infty$ four dimensional limit,
are given by the three roots of unity, in agreement with the
${\bf Z}_3$ symmetric set of singular points of the
Seiberg-Witten solution for $N\!=\!2$ four dimensional $SU(2)$
gauge theory, with $N_f=1$ \cite{SW2}. In fact, the minima of
(\ref{eq:32}) are at the points $\frac {1}{\tilde{x}}= \frac
{\Lambda e^{2 \pi i n /3}}{(2)^{1/3}}$, and therefore we get
\begin{equation}
\frac {\epsilon}{\tilde{x}} \simeq e^{2 \pi i n/3} \Lambda \epsilon, \: \: \: \: \: \: \:
\: n=0,1,2,
\label{eq:35}
\end{equation}
where $\gamma_1 \equiv \Lambda^3$, with $\Lambda$ the scale of
the $N\!=\!2$ theory. In the $\epsilon \rightarrow \infty$ limit,
the three ground states of the Coulomb branch approach each other
\cite{SW2}.

Using the mass deformed Atiyah-Hitchin manifold studied by
Dancer \cite{D},
\begin{equation}
y^2= x^2 v - i 2 m \gamma_1 x + \gamma_1^2, 
\label{eq:36}
\end{equation}
we can derive the superpotential for the massive case,
\begin{equation}
W= \epsilon \left( \gamma_1 \tilde{x} - \frac {2 i m}{\tilde{x}} +
\frac {1}{\tilde{x}^2} \right), 
\label{eq:37}
\end{equation}
which in the three dimensional $R \rightarrow 0$ limit leads to the superpotential 
\begin{equation} 
W= \epsilon \left( - \frac {2 i m}{\tilde{x}} +
\frac {1}{\tilde{x}^2} \right). 
\label{eq:38}
\end{equation}
  
Now, we could consider the type of non perturbative effects able
to generate, in the $N\!=\!2$ three dimensional theory with finite $\epsilon$, the superpotentials
(\ref{eq:33}) and (\ref{eq:38}). In the three dimensional
theory, we should use the Callias version of Atiyah-Hitchin
index theorem \cite{C}. In the notation of \cite{C}, the index is
given by 
\begin{equation}
\hbox{index} = [j(j+1)-\{m\}(\{m\}+1)] \cdot n,
\label{eq:39}
\end{equation}
for instanton number equal $n$. The $2n$ gluino zero modes
correspond to $j=1$ and $\{m\}=0$ in (\ref{eq:39}). For fermions
in the fundamental representation, $j=\frac {1}{2}$, and we get
$\hbox{index}=1$, for $\{m\}=- \frac {1}{2}$, and
$\hbox{index}=0$, for $\{m\}= \frac {1}{2}$\footnote{$\{m\}$ is
the largest eigenvalue of $\phi^{a}T^{a}$, with $\phi$ (the Higgs
field in the adjoint) smaller than the fermion mass.}. In order to
reproduce the mass term in (\ref{eq:37}), we should consider
$\{m\}=\frac {1}{2}$. The superpotential $\frac
{\epsilon}{\tilde{x}^2}$ is however more difficult to interpret.
According to the power of the denominator, we might think of some
sort of $2$-instanton effect. Taking into account the powers of
$\epsilon$, and interpretting the instanton effect as associated
to $\frac {\epsilon}{\tilde{x}}$, we get an instanton effect of
the type $\frac {\epsilon}{\tilde{x}}$, and other of the type
$\frac {\epsilon}{\tilde{x}} \cdot \left( \frac {1}{\epsilon}
\right)$. Thinking of this second instanton effect in similar
terms as the massive contribution $\frac {\epsilon}{\tilde{x}}
 (2im)$, we will get a net vertex $\lambda \lambda \lambda \lambda
\psi \psi$, which can generate the superpotential $\epsilon \frac
{1}{\tilde{x}^2}$, if we pair up $\lambda$ and $\psi$ zero modes,
and lift them. This is only an heuristic way to interpret the
quadratic term in (\ref{eq:32}) for finite values of $\epsilon$.
This effect is supressed in the $\epsilon \rightarrow \infty$
limit \cite{AHISS}. For finite $\epsilon$ that any singularity of
the four dimensional Coulomb branch leads to a confinement ground
state characterized by a vacuum expectation value for the
monopole fields. It would be interesting superpotentials of the
type (\ref{eq:32}), for finite $\epsilon$, with monopole
superpotentials.

%%%%%%%%%%%%%%%%%%%%%%%%%%%%%%%%%%%%%%%%%%%%%%%%%%%%%%%%%%%%%%%%%%
%%%%%%%%%%%%%%%%%%%%%%%%%%%%%%%%%%%%%%%%%%%%%%%%%%%%%%%%%%%%%%%%%%

For $N\!=\!1$ four dimensional $SU(2)$ gauge theory, with one massless 
flavor, we get in the instanton background six zero modes, four of them 
corresponding to superesymmetric and superconformal transformations 
acting on the instanton configuration, and the other two to 
the flavor fermionic zero modes. Using the technique of constrained 
instantons, it was shown in \cite{ADS} that a superpotential can be generated 
for $N_f\!=\!1$ by lifting four of the zero modes which can be paired. To 
derive this superpotential in the F-theory approach, we should consider 
the $D$-instanton defined fibering on $C$ the degenerate elliptic fiber 
of $A_1$ type. Again, taking into account the existence of an $SU(2)$ 
gauge connection on $C$, with topological number $1\!=\!N_f$, we get 
for this $D$ instanton aritmetic genus equal one \cite{BJPSV}. Notice that this four dimensional 
instanton has been directly 
defined in the F-theory context, where no resolution of the singularity is 
allowed.

%%%%%%%%%%%%%%%%%%%%%%%%%%%%%%%%%%%%%%%%%%%%%%%%%%%%%%%%%%%%%%%%%%%%%
%%%%%%%%%%%%%%%%%%%%%%%%%%%%%%%%%%%%%%%%%%%%%%%%%%%%%%%%%%%%%%%%%%%%% 

\section{Fractional Instantons.}

After the previous discussion on the superpotential 
a natural question arises, namely how to interpret the $R \rightarrow \infty$ 
limit of the superpotential (\ref{eq:11})  
directly in terms of topologically non trivial configurations in four 
dimensions. The more natural guess would be to think on some kind of 
``fractional instanton'', created by strong infrared dynamics: ordinary instantons in four dimensions 
posses too many zero modes to generate a superpotential. In fact, in the 
four dimensional F-theory context, where we can not perform any resolution 
of the singularity, the divisor $D$, with $\Pi(D)=C$, is of aritmetic  
genus $C_2(G)$. The fractional instanton should be thought as the M-theory 
$5$-brane, formally wrapping $D$ $\frac {1}{C_2(G)}$ times \cite{Wsp}. A different 
approach is the use of `t Hooft fractional instantons \cite{tHcmp} to generate 
the $R \rightarrow \infty$ limit of the superpotential (\ref{eq:11}). In the 
context of twisted boundary conditions on four dimensional spacetime \cite{tHtw}, 
a contribution in pure $N\!=\!1$ supersymmetric Yang-Mills to the superpotential 
can be expected from a ``toron'' (topological number $\frac {1}{N_C}$) 
amplitude \cite{CG}. Following the notation of reference
\cite{tHtw}, this amplitude in the infinite volume limit, 
will be given by 
\begin{equation}
<e,m=1| \lambda \lambda |e, m=1> = < m=1 | \lambda \lambda \: \: \Omega (k=1) | m=1> e^{2 \pi i
e/N_C}=\Lambda^3 e^{2 \pi i e /N_C}
\label{eq:23}
\end{equation}
where the electric flux $e$ can take values $e=0, \ldots, N_C-1$, and where 
the state $| m=1 >$ corresponds, in the temporal gauge $A^0=0$, to a 
fractional magnetic flux configuration, of magnetic flux $\frac {1}{N_C}$ in the third direction. The 
gauge transformation $\Omega (k=1)$ is part of the residual gauge 
symmetry, compatible with the twisted boundary conditions corresponding 
to the existence of the magnetic flux $|m=1>$. The phase factor in (\ref{eq:23}) 
comes from the definition of invariant states with respect to this residual 
symmetry:
\begin{equation}
|\overline{e},\overline{m}> \equiv  \frac {1}{N_C^3} \sum_{k} e^{2 \pi i \overline{k} 
\overline{e}/N_C} \Omega(k) |\overline{m}>.
\label{eq:24}
\end{equation}
  
The fractional Pontryagin number corresponding to the tunnelling amplitude 
in (\ref{eq:23}) is given by
\begin{equation}
P = \frac {1}{N_C}.
\label{eq:25}
\end{equation}
  
Now, it is easy to observe that 
\begin{equation}
\Omega (k=1) ^{N_C} = T,
\label{eq:26}
\end{equation}
where $T$ is a periodic gauge transformation on $S^3$, with $\Pi_3=1$. Taking into 
account that the non abelian instanton is defined by fibering the elliptic 
fiber $E$ on $C$, relation (\ref{eq:26}) becomes the analog of the exponentiated 
version of Dynkin 
relation (\ref{eq:9}) for $SU(N_C)$. Defining the instanton action as $\hbox{exp} - \frac {1}{g_4^2}$, 
the sum of contributions of type (\ref{eq:23}) to the superpotential produces, 
by means of (\ref{eq:26}), the desired result (\ref{eq:11}). Moreover, the 
minima for $SU(N_C)$ of the superpotential (\ref{eq:11}) reproduce exactly 
the phases in (\ref{eq:23}), which come from the definition of electric 
flux through (\ref{eq:24}). Notice that the minima of (\ref{eq:11}) 
are parametrized by the Wilson loop in the internal direction, which is precisely the meaning 
of $e$ in (\ref{eq:23}). The fact that $<\lambda \lambda>$ is given in four dimensions in 
terms of $e$, comes in the toron computation directly from the the definition (\ref{eq:24}) 
of invariant states. Thus, we conclude that fractional instantons effectively appear 
in the uncompactified four dimensional limit. Notice that the Wilson loop in the internal 
direction, which parametrizes the Coulomb branch, can take values, if flavors are absent, 
in the center ${\bf Z}_{N_C}$ of the color group even when gauge invariance is restored. 
The discrete set
of points in the moduli corresponding to Wilson lines in the center, 
becomes the ground states in the uncompactified limit. 
At these points, we have 
vortices instead of monopoles and torons as their twisted version, that survive in the four
dimensional limit. The torons appearing as the relevant topological configurations at these points.
 The crucial
dynamics of torons that provide the right counting of zero modes
consists in avoiding the superconformal zero mode modes of the
instanton, without breaking supersymmetry; furthermore in the infinite volume limit they have
vanishinjg field strength. In summary, we believe
that fractional instanton effects in four dimensional $N\!=\!1$ are the right ingredient to understand the 
superpotentials derived through M-theory techniques, in the uncompactified limit.

\vspace{2 cm}
  
\begin{center}
  
{\bf Acknowledgements}
  
\end{center}
  
This research is partially supported under grant AEN $96$-$1655$, and 
by E. C. grant FMRX-CT $960012$.

\end{document}